\documentclass[prd,showpacs,preprintnumbers,twocolumn,amsmath,nofootinbib,amssymb]{revtex4}
\usepackage{graphicx,color,dcolumn,booktabs,bm}
\usepackage{epstopdf}
\usepackage{longtable,lscape}
\usepackage{txfonts}
\usepackage{overpic}
\usepackage{float}
\usepackage{amssymb}
\usepackage{amsmath}
\usepackage{epstopdf}
\usepackage{indentfirst}
\usepackage{feynmf}   
\usepackage{slashed}  
\usepackage{cases}
\usepackage{ulem}
\usepackage{color}
\usepackage{float}
\usepackage{multirow}
\usepackage{tikz}
\usepackage{epstopdf}
\usepackage{epsfig,dsfont,amssymb,amsmath,amsfonts,amsbsy,mathrsfs}
\usepackage{multirow}
\usepackage{graphicx}
\usepackage{float}
\usepackage{subfigure}
\usepackage{array}

\usepackage[colorlinks, citecolor=blue,anchorcolor=red,menucolor=red, linkcolor=red,filecolor=red,runcolor=red,urlcolor=blue,frenchlinks=red]{hyperref}
\makeatletter
\@addtoreset{equation}{section}
\makeatother

\newcommand{\nc}{\newcommand}
\nc{\tj}[1]{\textcolor{red}{Tianjie: #1}}
\allowdisplaybreaks
\begin{document}
\title{Study of the $\omega$ and $\omega_3$, $\rho$ and $\rho_3$ and the newly observed $\omega$-like state $X(2220)$}
\author{Ya-Rong Wang$^{1,2}$}\email{nanoshine@foxmail.com}
\author{Ting-Yan Li$^{1,2}$}\email{litingyan1213@163.com}
\author{Zheng-Yuan Fang$^{1,2}$}\email{fang1628671420@163.com}
\author{Hao Chen$^{1,2,3}$}\email{chenhao0602@live.cn}
\author{Cheng-Qun Pang$^{1,2,3}$\footnote{Corresponding author}}\email{xuehua45@163.com}
\affiliation{$^1$College of Physics and Electronic Information Engineering, Qinghai Normal University, Xining 810000, China\\$^2$Joint Research Center for Physics,
Lanzhou University and Qinghai Normal University,
Xining 810000, China \\$^3$Lanzhou Center for Theoretical Physics, Key Laboratory of Theoretical Physics of Gansu Province, Lanzhou University, Lanzhou, Gansu 730000, China}

\begin{abstract}

  We study the excited states of $\omega$ and $\omega_3$ by comparison  with the $\rho$ and $\rho_3$  families, and discuss the possibility of $X(2220)$ as $\omega$ excitation by analyzing the mass spectra and strong decay behaviors. In addition, we predict the masses and widths of $\omega(2D)$ and $\omega_3$and $\rho_3(4D)$, $\rho_3(1G)$, $\omega_3$ and $\rho_3(2G)$ and $\omega_3(3G)$ and $\rho_3(3G)$. The abundant information of their two-body strong decays predicted in this work will be helpful to further study of these $\omega$ and $\omega_3$ and $\rho$ and $\rho_3$ states in experiment and theory.

\end{abstract}
\maketitle

\section{introduction}\label{sec1}

Very recently, the BESIII Collaboration announced the observation of the $X(2220)$ state in the $e ^+e^- \to \omega\pi^0\pi^0$ process \cite{BESIII:2021uni}, which has a statistical significance larger than $5\sigma$.
The measurement shows that the $X(2220)$ has the mass of $M = 2222 \pm 7 \pm 2$ MeV and the width of $\Gamma = 59 \pm 30 \pm 6$ MeV.
It could be an $\omega$ excited state from its report.
The Lanzhou group indicates that the enhancement structures around 2.2 GeV existing in $e^+e^- \to \omega\eta$ and $e^+e^- \to \omega\pi^0\pi^0$ contain the $\omega(4S)$ and $\omega(3D)$ signals \cite{Zhou:2022wwk}.
The BESIII Collaboration announced $X(2220)$ could be an excited $\omega$ resonance \cite{BESIII:2022mxl}.
Present studies show that it is reasonable to regard $X(2220)$ as an $\omega$ state.
When further checking the experimental status of light mesons, we notice an interesting phenomenon: the states of unflavored light meson families with  $3^{--}$ are not established well. In our previous work \cite{Wang:2019jch}, we carried out a system investigation of the $\omega$ mesons and studied the $\omega$-like $X(2240)$ state observed by BESIII Collaboration \cite{BESIII:2018ldc}. Now, we reconsider the newly observed $X(2220)$ state and the $\omega$ and $\omega_3$, $\rho$ and $\rho_3$ families.

As an important part of the meson family, the $\omega$ and $\rho$ and $\omega_3$ and $\rho_3$ families have become more and more abundant in experiment. BESIII Collaboration reported a result that in the $e^+e^- \to \omega\pi^0$ cross section, a resonance denoted by $Y(2040)$ was observed  recently. Its mass and width are determined to be $2034 \pm 13 \pm 9$ and $234 \pm 30 \pm 25 $ MeV \cite{Ablikim:2020das}, respectively; Ref. \cite{Li:2021qgz} also treated it as $\rho(2000)$. BESIII Collaboration also reported a state []we call it $X(2239)$] with the mass of $M=2239.2\pm7.1\pm11.3$ MeV and the width of  $\Gamma=139.8\pm12.3\pm20.6$ MeV \cite{Ablikim:2018iyx} by studying the cross section of the $e^+e^-\to K^+K^-$ process. $X(2239)$ is treated as $Y(2175)$  in the Ref. \cite{Chen:2020xho}. Reference \cite{Wang:2019jch} assigned $X(2239)$ as the $\omega(4^3S_1)$ state and it treated $X(2239)$ as the candidate of tetraquark states in Refs. \cite{Lu:2019ira,Azizi:2019ecm}. Barnes, Godfrey, and Swanson suggest that further investigation is expected to understand the structures near the $\Lambda\bar\Lambda$ threshold, such as $X(2239)$, $\eta(2225)$, and $X(2175)$. However, the properties of $X(2240)$ still remain unclear \cite{Zhu:2019ibc}.

Afonin classified light nonstrange mesons according to the values of $(L, n)$ and indicated that $\omega_3(1670)$, $\omega_3(2285)$, $\rho_3(1690)$, are $D$-wave mesons \cite{Afonin:2007aa}.
 In 2009, Ebert calculated masses of the ground state as well as the orbital and radial excited states of quark-antiquark mesons within the relativistic quark model, which included the masses of $\omega$, $\omega_3$, $\rho$ and $\rho_3$ meson families \cite{Ebert:2009ub}.
Anisovish $et~al$. consider the decays of $\omega$ and $\rho$ mensons, the $^3S_1~q\bar{q}$ states of constituent quark model: $\rho \to \gamma\pi$, $\omega \to \gamma\pi$, and $\rho \to \pi\pi$ and give correct values for the partial widths of radiative and hadronic decays of confined $q\bar{q}$ states \cite{Anisovich:2010zza}.
In 2013, the Lanzhou group studied $\rho$, $\rho_3$ family \cite{He:2013ttg}, which helps to construct the whole $\rho$ family. Later, Wang $et~al$. predicted the spectroscopy behavior of $\omega$, $\rho$, and $\phi$ mesons at the range of $2.4 \sim 3$ GeV \cite{Wang:2021abg}.
Feng $et~al$. analyze the masses and two-body strong decay widths of $^3S_1$ vector mesons \cite{feng:2021igh}.
The masses and strong decay widths of the higher excited $\rho$ mesons are calculated in Ref. \cite{Feng:2022hwq}.
All of the above mentioned works refer to $\omega$ and $\omega_3$ and $\rho$ and $\rho_3$ mesons.

Not only are the lightest vector meson resonances, $\rho$, $\omega$, and $\phi$  benchmark states in understanding of the quark substructure of hadrons \cite{Johnson:2020ilc}, but also they are important parts of the light meson family.
A systematic study for the $\omega$, $\rho$, and $\omega_3$, $\rho_3$ meson family  becomes very necessary and urgent. $\omega$ and $\omega_3$ and $\rho$ and $\rho_3$ mesons have same quantum numbers ($P$ and $C$), for which reason the study of them can be borrowed from each other. We estimate the mass of the missing states in these meson families. Similarly, we study the strong decays of the $\omega$ and $\omega_3$ meson family comparing with that of the $\rho$ and $\rho_3$ meson family.

In this work, we will study the excited states of $\omega$, $\rho$, and $\omega_3$, $\rho_3$ mesons. By using the modified Godfrey-Isgur quark (MGI) model and quark pair creation (QPC) model, the mass spectra and strong decay behavior of the excited states of these mesons are analyzed, which indicates  that $X(2220)$  is the candidate of $\omega(3D)$ meson with $I(J^P)=0(1^-)$. The spectra of the $\omega$, $\rho$ and $\omega_3$, $\rho_3$ meson families are studied.

This paper is organized as follows. We first introduce the models applied in this paper in section \ref{sec2} and analyze the mass spectra and decay behaviors of the $\omega, \rho$ and $\omega_3, \rho_3$  mesons  in section \ref{sec3}. The paper ends with a summary.

\section{Models employed in the work} \label{sec2}

The modified GI quark model, Regge trajectory, and the QPC model can be used to calculate the mass spectra and the two-body strong decays of the  meson family, respectively. Before discussing the results, the above models will be introduced briefly here.

\subsection{Regge trajectory}
It is efficient to investigate a light-meson spectrum by determining the Regge trajectory \cite{Chew:1962eu,Anisovich:2000kxa}.
 The following relation is satisfied by the masses and radial quantum numbers of the light mesons that belong to the same meson family:
\begin{eqnarray}
M^2=M_0^2+(n-1)\mu^2, \label{rt}
\end{eqnarray}
where {$M_0$} represents the mass of the ground state, $n$ is the radial quantum number of the corresponding meson with  the mass $M$, and $\mu^2$ is the trajectory slope.
 The Regge trajectories are a great method to help us decide the approximate mass range of a particle and it lays a good foundation for our next step.
This method can approximately hold even if it is not as strict as the Regge trajectory as a function of the particle spin, it can still approximately hold since at most one or two excitations are known for each quantum number. And we will calculate the mass spectrum using the MGI model, which obviously takes into account spin breaking.

\begin{figure*}[htbp]
\centering%
\includegraphics[height=2in, width=6.5in]{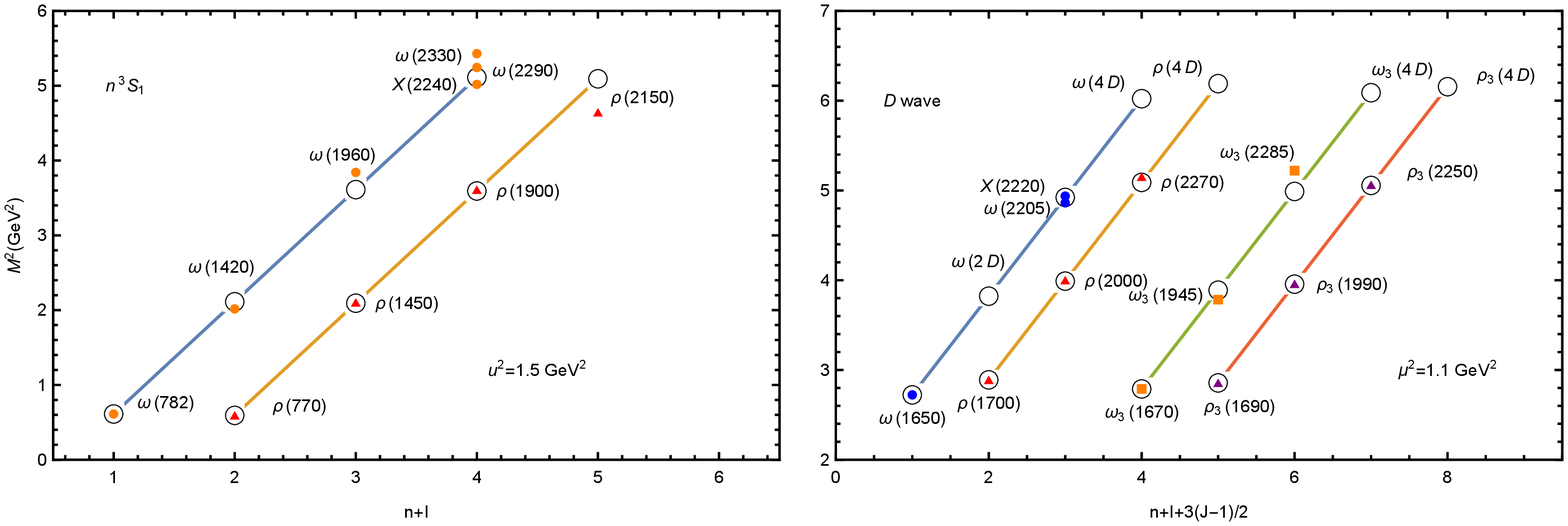}
\caption{The $S$ and $D$ waves' Regge trajectories of $\omega$, $\omega_3$ and $\rho$, $\rho_3$
families. Here $J$ is the angular momentum of the meson. The open circle and the filled geometry are the theoretical and experimental values, respectively.}
\label{regge}
\end{figure*}


\subsection{The modified GI model}
Godfrey and Isgur successfully proposed the GI model in 1985 for characterizing relativistic meson spectra, particularly for low-lying mesons \cite{Godfrey:1985xj}.
In order to better describe the spectrum of the highly excited states, Song $et~al$. \cite{Song:2015nia,Song:2015fha} introduced the color screening effect on the basis of the Cornell potential to modify the GI model (MGI model).

The Hamiltonian of the potential model reads
\begin{equation}\label{Hamtn}
  \tilde{H}=\sqrt{m_1^2+\mathbf{p}^2}+\sqrt{m_2^2+\mathbf{p}^2}+\tilde{V}_{\mathrm{eff}}(\mathbf{p,r}),
\end{equation}
which represents the internal interactions of mesons.
$m_1$ and $m_2$ are the mass of quark and antiquark, respectively, the
 effective potential has a familiar format in the nonrelativistic limit before smeared  \cite{Godfrey:1985xj,Lucha:1991vn}
\begin{eqnarray}
V_{\mathrm{eff}}(r)=H^{\mathrm{conf}}+H^{\mathrm{hyp}}+H^{\mathrm{so}}\label{1},
\end{eqnarray}
with
\begin{align}
 H^{\mathrm{conf}}&=\Big[-\frac{3}{4}(\frac{b(1-e^{-\mu r})}{\mu}+c)+\frac{\alpha_s(r)}{r}\Big](\bm{F}_1\cdot\bm{F}_2)\nonumber\\ &=S(r)+G(r),\label{3}\\
H^{\mathrm{hyp}}&=-\frac{\alpha_s(r)}{m_{1}m_{2}}\Bigg[\frac{8\pi}{3}\bm{S}_1\cdot\bm{S}_2\delta^3 (\bm r) +\frac{1}{r^3}\Big(\frac{3\bm{S}_1\cdot\bm r \bm{S}_2\cdot\bm r}{r^2} \nonumber  \\ \label{3.1}
&\quad -\bm{S}_1\cdot\bm{S}_2\Big)\Bigg] (\bm{F}_1\cdot\bm{F}_2), \\
H^{\mathrm{so}}=&H^{\mathrm{so(cm)}}+H^{\mathrm{so(tp)}},  \label{3.2}
\end{align}
where $\bm{S}$ indicates the spin of quark or antiquark and $\bm{L}$ is the orbital momentum.  For a meson, $\langle\bm{F}_1\cdot\bm{F}_2 \rangle=-4/3$, the running coupling constant $\alpha_s(r)$ has the following form:

 \begin{align}
  \alpha_s(r)=\sum_{k=1}^3\frac{2\alpha_k}{\sqrt{\pi}} \int_{0}^{\gamma_k r}e^{-x^2}dx,
 \end{align}
 where $\alpha_{1,2,3}=0.25,0.15,0.2$  and $\gamma_{1.2.3}=\frac{1}{2},\frac{\sqrt{10}}{2},\frac{\sqrt{1000}}{2}$ \cite{Godfrey:1985xj}.
$H^{\mathrm{conf}}$ reflects the spin-independent interaction, and it can divided into two parts: $S(r)$ and $G(r)$.
$H^{\mathrm{hyp}}$ represents the color-hyperfine interaction. $H^{\mathrm{so}}$ is the spin-orbit interaction that contains the color magnetic term and the Thomas precession term. $H^{\mathrm{so}}$ can be written as

\begin{eqnarray}
H^{\mathrm{so(cm)}}=\frac{-\alpha_s(r)}{r^3}\left(\frac{1}{m_{1}}+\frac{1}{m_{2}}\right)\left(\frac{\bm{S}_1}{m_{1}}+\frac{\bm{S}_2}{m_{2}}\right)
\cdot
\bm{L}(\bm{F}_1\cdot\bm{F}_2),
\end{eqnarray}
\begin{eqnarray}
H^{\mathrm{so(tp)}}=-\frac{1}{2r}\frac{\partial H^{\mathrm{conf}}}{\partial
r}\Bigg(\frac{\bm{S}_1}{m^2_{1}}+\frac{\bm{S}_2}{m^2_{2}}\Bigg)\cdot \bm{L}.
\end{eqnarray}
The parameters of  the MGI model are defined in Table \ref{SGIfit1} \cite{Pang:2018gcn}.

\begin{table}[htbp]
\caption{Parameters and their values in this work, which are given by Ref. \cite{Pang:2018gcn}. \label{SGIfit1}}

\begin{center}
\begin{tabular}{cccc}
\toprule[1pt]\toprule[1pt]
Parameter &  value \cite{Pang:2018gcn}&Parameter &  value \cite{Pang:2018gcn} \\
 \midrule[1pt]
$m_u$ (GeV) &0.163&{$\sigma_0$ (GeV)}&{1.799}\\
$m_d$ (GeV) &0.163&{$s$ (GeV)}&{1.497}\\
$m_s$ (GeV) &0.387&$\mu$ (GeV)&0.0635 \\
$b$ (GeV$^2$) &0.221&$c$ (GeV)&-0.240\\
$\epsilon_c$&-0.138&$\epsilon_{sov}$&0.157\\
$\epsilon_{sos}$&0.9726& $\epsilon_t$&0.893\\
\bottomrule[1pt]\bottomrule[1pt]
\end{tabular}
\end{center}
\end{table}

Through diagonalizing and solving the Hamiltonian in Eq. (\ref{Hamtn}) by exploiting a simple harmonic oscillator (SHO) basis, the mass spectra and wave functions can be obtained. The mass and wave function of the meson are applied to the strong decay process.

\subsection{A brief review of QPC mdoel}
The QPC model is used to calculate the  hadronic strong decays allowed by the Okubo-Zweig-Iizuka(OZI) rule. This model was initially put up by Micu \cite{Micu:1968mk} and further developed by the Orsay group \cite{LeYaouanc:1972ae, LeYaouanc:1973xz, LeYaouanc:1974mr, LeYaouanc:1977gm, LeYaouanc:1977ux}.
The QPC model is widely applied to the OZI-allowed two-body strong decays  of mesons in Refs. \cite{vanBeveren:1979bd, vanBeveren:1982qb, Capstick:1993kb, Page:1995rh, Titov:1995si, Ackleh:1996yt, Blundell:1996as,
Bonnaz:2001aj, Zhou:2004mw, Lu:2006ry, Zhang:2006yj, Luo:2009wu, Sun:2009tg, Liu:2009fe, Sun:2010pg, Rijken:2010zza, Ye:2012gu,
Wang:2012wa, He:2013ttg, Sun:2013qca, Pang:2014laa, Wang:2014sea, Chen:2015iqa, Pang:2017dlw, Pang:2018gcn}.
The transition operator $\mathcal{T}$ describes a quark-antiquark pair (denoted by indices $3$ and $4$)  creation from vacuum that
$J^{PC}=0^{++}$.
For the process $A\to B+C$, $\mathcal{T}$ can be written as \cite{Wang:2019jch}
{\begin{align}\label{gamma}
\mathcal{T} = & -3\gamma \sum_{m}\langle 1m;1~-m|00\rangle\int d \mathbf{p}_3d\mathbf{p}_4\delta ^3 (\mathbf{p}_3+\mathbf{p}_4) \nonumber \\
 & ~\times \mathcal{Y}_{1m}\left(\frac{\textbf{p}_3-\mathbf{p}_4}{2}\right)\chi _{1,-m}^{34}\phi _{0}^{34}
\left(\omega_{0}^{34}\right)_{ij}b_{3i}^{\dag}(\mathbf{p}_3)d_{4j}^{\dag}(\mathbf{p}_4),
\end{align}}
where $\mathcal{Y}_{\ell m}(\mathbf{p})={|\mathbf{p}|^{\ell}}Y_{\ell
m}(\mathbf{p})$ represents the solid harmonics. $\chi$, $\phi$, and $\omega$ denote the spin, flavor, and color wave functions respectively. Subindices $i$ and $j$ are the color index of a $q\bar{q}$ pair. The amplitude ${M}^{{M}_{J_{A}}M_{J_{B}}M_{J_{C}}}$ of the decay process is defined with the transition operator $T$,

\begin{eqnarray}
\langle BC|\mathcal{T}|A \rangle = \delta ^3(\mathbf{P}_B+\mathbf{P}_C)\mathcal{M}^{{M}_{J_{A}}M_{J_{B}}M_{J_{C}}},
\end{eqnarray}
$\mathbf{P}$ is the three-momentum  in the rest frame of a meson A.
Finally, the total width can be expressed in terms of the partial wave amplitude squared,
\begin{eqnarray}
\Gamma&=&\frac{\pi}{4} \frac{|\mathbf{P}|}{m_A^2}\sum_{J,L}|\mathcal{M}^{JL}(\mathbf{P})|^2,
\end{eqnarray}
where $m_{A}$ is the mass of an initial state $A$, and
the two decay amplitudes can be related by the Jacob-Wick formula \cite{Jacob:1959at} as

\begin{equation}
\begin{aligned}
\mathcal{M}^{JL}(\mathbf{P}) = &\frac{\sqrt{4\pi(2L+1)}}{2J_A+1}\sum_{M_{J_B}M_{J_C}}\langle L0;JM_{J_A}|J_AM_{J_A}\rangle \\
    &\times \langle J_BM_{J_B};J_CM_{J_C}|{J_A}M_{J_A}\rangle \mathcal{M}^{M_{J_{A}}M_{J_B}M_{J_C}}.
\end{aligned}	
\end{equation}

\section{Numerical results and phenomenological analysis}\label{sec3}
We adopt the Regge trajectory and the MGI model to calculate the mass spectrums of $\omega$ and $\omega_3$ and $\rho$ and $\rho_3$ mesons and employ the QPC model
with the meson wave functions obtained from
the MGI model to evaluate the meson' decay widths.
 Following, we will give the numerical results and phenomenological analysis.

\subsection{Mass spectra analysis}

 We plot the Regge trajectory of  $\omega$ and $\omega_3$, $\rho$ and $\rho_3$ meson families in the Fig. \ref{regge}. By analyzing the Regge trajectory, the results demonstrate that $\omega(782)$, $\omega(1420)$, and $\omega(1960)$ are $1S$, $2S$, and $3S$ states, respectively. $\rho(770)$, $\rho(1450)$, $\rho(1900)$, and $\rho(2150)$ are $1S$, $2S$, $3S$, and $4S$ states.
However, the situation of $\omega(4S)$ is not so distinct. $\omega(2330)$, $\omega(2290)$, and $X(2240)$ can be the candidates of $\omega(4S)$ because of their approximate masses. The widths of those states are different. The mass of $\omega(2330)$ is larger than that of others, more details can be seen in the Fig. \ref{regge} and the numerical result in Table \ref{mass}.
As for the $D$ wave, the states $\rho(1700)$, $\rho(2000)$, and $\rho(2270)$ are on the ($n + I + J, M^2$) planes. We assign $X(2220)$ and $\omega(2205)$ as the $3^3D_1$ states. $\omega_{3}(1945)$ and $\omega_{3}(2285)$ are the excited states of $\omega_3(1670)$. $\rho_3(1990)$ and $\rho_3(2250)$ are the $2D$ and $3D$ states, which agrees well with the results in Ref. \cite{He:2013ttg}. We can see the masses of $\omega(1650)$ and $\rho(1700)$, $\omega_3(1670)$ and $\rho_3(1670)$, $\omega_3(1945)$ and $\rho_3(1990)$ are very similar.

Applying the MGI model and the parameters in Table \ref{SGIfit1}, the mass spectra of $S$ wave and $D$ wave of $\omega$ and $\rho$ mesons as well as $\omega_3$ and $\rho_3$ mesons are shown in the Table \ref{mass}.
Jia $et~al$. revisited the orbital Regge spectra of the light unflavored mesons \cite{Jia_2017}. And as we reviewed in section \ref{sec1}, Anisovich $et~al$. investigated the Regge trajectory   \cite{2005Quark}, Ebert $et~al$. gave the mass spectrum of the $\omega$ and $\omega_3$ and $\rho$ and $\rho_3$ mesons \cite{Ebert:2009ub}. Our numerical results are compared with experimental masses and their theoretical results  in Table \ref{mass}.
As shown in Table \ref{mass}, Ref. \cite{2005Quark} estimated the mass of $\omega_3(1D)$ to be 1671 MeV, which is smaller than the mass obtained with the MGI model and Ref. \cite{Ebert:2009ub}. The mass of $\omega(1^3S_1)$ calculated in the MGI model is smaller than that in Refs. \cite{Jia_2017,2005Quark,Ebert:2009ub}. The mass spectrum of Ref. \cite{2005Quark} and MGI model all indicate that the newly observed state $X(2220)$ may be an $\omega(3D)$ state.
Furthermore, we find that the lower energy excited states such as $1S$, $2S$ and $1D$, $2D$ are consistent with experimental values \cite{Zyla:2020zbs, bugg2004four}.

As for $\omega_3$ and $\rho_3$ meson families, there is too little information to help us determine their structures. For the states of $D$ wave, $\omega_3(1670)$ and $\rho_3(1690)$ are the $1D$ states which has been certified by many researchers \cite{Buisseret:2004wm, Ebert:2009ub, Afonin:2007aa, Pang:2015eha}. With respect to $2D$ states, we predict the masses of $\omega(2D)$ and $\rho(2D)$ are both $2074$ MeV.  At the same time, $\rho(2000)$  is a $\rho(2D)$ state \cite{He:2013ttg, Afonin:2007aa}, which is also close to our theoretical prediction. With regard to the $3D$ states of $\omega_3$ and $\rho_3$, the value of the mass that we calculated is $2376.5$ MeV. Moreover, we consider $\rho_3(2250)$ as the $\rho_3(3D)$ state, which agrees well with the result in Ref. \cite{He:2013ttg}.

Besides, we have also calculated the mass spectra of $\omega_3$ and $\rho_3$ of $G$ wave. The masses of $1G$, $2G$, and $3G$ are $2254.8$, $2522.3$ MeV, and $2748.5$, respectively. Further discussion based on the decay behaviors will be given in the next section.  We hope that this information can help find these resonances in experiment.

\begin{table*}[htbp]
\caption{The mass spectra of $\omega$ and $\omega_3$ and $\rho$ and $\rho_3$  mesons.
The unit is MeV. \label{mass}}
\vspace{-0.8cm}
\begin{center}
\[\begin{array}{cccccc}
\hline
\hline
n^{2s+1}L_J   &\text{Jia} $\cite{Jia_2017}$    &\text{Anisovish} $\cite{2005Quark}$
&\text{Ebert} $\cite{Ebert:2009ub}$  &\text{This~work}         &\text{Experimental mass}\\
\omega(1^3S_1)  &{846}   & \multicolumn{1}{c}{\multirow{2}{*}{778}} & \multicolumn{1}{c}{\multirow{2}{*}{776}} & \multicolumn{1}{c}{\multirow{2}{*}{773.8}} &\multicolumn{1}{c}{$ 782.65$ \pm $0.12~\cite{Zyla:2020zbs}$}\\
\rho(1^3S_1)   &775 &\multicolumn{1}{r}{}   & \multicolumn{1}{r}{}&\multicolumn{1}{r}{}&\multicolumn{1}{c}{$775.26$ \pm  $0.25 \cite{Zyla:2020zbs}$} \\
\omega(2^3S_1)  &{ }   & \multicolumn{1}{c}{\multirow{2}{*}{1473}} & \multicolumn{1}{c}{\multirow{2}{*}{1486}} & \multicolumn{1}{c}{\multirow{2}{*}{1424.3}} &\multicolumn{1}{c}{$1410$ \pm$ 60 \cite{Zyla:2020zbs}$}\\
\rho(2^3S_1)   &   &\multicolumn{1}{c}{ }   & \multicolumn{1}{c}{ }&\multicolumn{1}{c}{ }&\multicolumn{1}{c}{$1465$ \pm $25~\cite{Zyla:2020zbs}$} \\
\omega(3^3S_1)  &{ }   & \multicolumn{1}{c}{\multirow{2}{*}{1763}} & \multicolumn{1}{c}{\multirow{2}{*}{1921}} & \multicolumn{1}{c}{\multirow{2}{*}{1906.1}} &\multicolumn{1}{c}{$1960$ \pm $25 \cite{Zyla:2020zbs}$}\\
\rho(3^3S_1)   &   &\multicolumn{1}{c}{ }   & \multicolumn{1}{c}{ }&\multicolumn{1}{c}{ }&\multicolumn{1}{c}{$1909 $ \pm $ 30.2 \cite{aubert2008measurements}$} \\
\omega(4^3S_1) &   &\multicolumn{1}{c}{\multirow{2}{*}{2158}}
& \multicolumn{1}{c}{\multirow{2}{*}{2195}}  &\multicolumn{1}{c}{\multirow{2}{*}{2259.1}} &\multicolumn{1}{c}{$2330 $ \pm $ 30 \cite{Zyla:2020zbs}$, $2290 $ \pm $ 20 \cite{Zyla:2020zbs}$, $2239.2 $ \pm $ 7.1 $ \pm $ 11.3 \cite{Ablikim:2018iyx}$}\\
\rho(4^3S_1)   &   &\multicolumn{1}{c}{ }   & \multicolumn{1}{c}{ }&\multicolumn{1}{c}{ }&\multicolumn{1}{c}{2201 \pm $19 \cite{PhysRevD.101.012011}$} \\
\omega(5^3S_1) &   &\multicolumn{1}{c}{\multirow{2}{*}{2363}}
& \multicolumn{1}{c}{\multirow{2}{*}{ }}  &\multicolumn{1}{c}{\multirow{2}{*}{2542.1}} &\multicolumn{1}{c}{-- }\\
\rho(5^3S_1)   &   &\multicolumn{1}{c}{ }   & \multicolumn{1}{c}{ }&\multicolumn{1}{c}{ }&\multicolumn{1}{c}{-- } \\\hline
\omega(1^3D_1) &   &\multicolumn{1}{c}{\multirow{2}{*}{1701}}
& \multicolumn{1}{c}{\multirow{2}{*}{1557}}  &\multicolumn{1}{c}{\multirow{2}{*}{1646.2}} &\multicolumn{1}{c}{$1670 $\pm$ 30 \cite{Zyla:2020zbs}$}\\
\rho(1^3D_1)   &   &\multicolumn{1}{c}{ }   & \multicolumn{1}{c}{ }&\multicolumn{1}{c}{ }&\multicolumn{1}{c}{ $1720 $\pm$ 20 \cite{Zyla:2020zbs}$} \\
\omega(2^3D_1) &   &\multicolumn{1}{c}{\multirow{2}{*}{1992}}
& \multicolumn{1}{c}{\multirow{2}{*}{1895}}  &\multicolumn{1}{c}{\multirow{2}{*}{2047.6}} &\multicolumn{1}{c}{--}\\
\rho(2^3D_1)   &   &\multicolumn{1}{c}{ }   & \multicolumn{1}{c}{ }&\multicolumn{1}{c}{ }&\multicolumn{1}{c}{$2000 $\pm$ 30 \cite{bugg2004four}$} \\
\omega(3^3D_1) &   &\multicolumn{1}{c}{\multirow{2}{*}{2212}}
& \multicolumn{1}{c}{\multirow{2}{*}{2168}}  &\multicolumn{1}{c}{\multirow{2}{*}{2365.0}} &\multicolumn{1}{c}{$2205 $\pm$ 30 \cite{Zyla:2020zbs}$, $2222 $\pm$ 7.3 \cite{BESIII:2021uni}$ }\\
\rho(3^3D_1)   &   &\multicolumn{1}{c}{ }   & \multicolumn{1}{c}{ }&\multicolumn{1}{c}{ }&\multicolumn{1}{c}{$2265 $\pm$ 40 \cite{Zyla:2020zbs}$ } \\

\omega(4^3D_1) &   &\multicolumn{1}{c}{\multirow{2}{*}{ }}
& \multicolumn{1}{c}{\multirow{2}{*}{ }}  &\multicolumn{1}{c}{\multirow{2}{*}{2624.0}} &\multicolumn{1}{c}{--}\\
\rho(4^3D_1)   &   &\multicolumn{1}{c}{ }   & \multicolumn{1}{c}{ }&\multicolumn{1}{c}{ }&\multicolumn{1}{c}{--} \\\hline
\omega_3(1^3D_3) &1661   &\multicolumn{1}{c}{\multirow{2}{*}{1671}}
& \multicolumn{1}{c}{\multirow{2}{*}{1714}}  &\multicolumn{1}{c}{\multirow{2}{*}{1708.2}} &\multicolumn{1}{c}{$1667$\pm$4 \cite{Zyla:2020zbs}$}\\
\rho_3(1^3D_3)   &1675   &\multicolumn{1}{c}{ }   & \multicolumn{1}{c}{}&\multicolumn{1}{c}{ }&\multicolumn{1}{c}{$1688.8$\pm$2.1 \cite{Zyla:2020zbs}$} \\
\omega_3(2^3D_3) &    &\multicolumn{1}{c}{\multirow{2}{*}{1987}}
& \multicolumn{1}{c}{\multirow{2}{*}{2066}}  &\multicolumn{1}{c}{\multirow{2}{*}{2074.0}} &\multicolumn{1}{c}{$1945$\pm$20 \cite{Zyla:2020zbs}$ }\\
\rho_3(2^3D_3)   &    &\multicolumn{1}{c}{ }   & \multicolumn{1}{c}{}&\multicolumn{1}{c}{ }&\multicolumn{1}{c}{$1982$\pm$14 \cite{Zyla:2020zbs}$} \\
\omega_3(3^3D_3) &    &\multicolumn{1}{c}{\multirow{2}{*}{2376}}
& \multicolumn{1}{c}{\multirow{2}{*}{2309}}  &\multicolumn{1}{c}{\multirow{2}{*}{2374.0}} &\multicolumn{1}{c}{$2285$\pm$60 \cite{Zyla:2020zbs}$ }\\
\rho_3(3^3D_3)   &    &\multicolumn{1}{c}{ }   & \multicolumn{1}{c}{}&\multicolumn{1}{c}{ }&\multicolumn{1}{c}{$2232 \cite{Zyla:2020zbs}$} \\
\omega_3(4^3D_3) &    &\multicolumn{1}{c}{\multirow{2}{*}{2705}}
& \multicolumn{1}{c}{\multirow{2}{*}{ }}  &\multicolumn{1}{c}{\multirow{2}{*}{2628.8}} &\multicolumn{1}{c}{--}\\
\rho_3(4^3D_3)   &    &\multicolumn{1}{c}{ }   & \multicolumn{1}{c}{}&\multicolumn{1}{c}{ }&\multicolumn{1}{c}{--} \\\hline
\omega_3(1^3G_3) &    &\multicolumn{1}{c}{\multirow{2}{*}{2252}}
& \multicolumn{1}{c}{\multirow{2}{*}{2002}}  &\multicolumn{1}{c}{\multirow{2}{*}{2254.8}} &\multicolumn{1}{c}{$2255$\pm$15 \cite{Zyla:2020zbs}$}\\
\rho_3(1^3G_3)   &    &\multicolumn{1}{c}{ }   & \multicolumn{1}{c}{}&\multicolumn{1}{c}{ }&\multicolumn{1}{c}{--} \\
\omega_3(2^3G_3) &    &\multicolumn{1}{c}{\multirow{2}{*}{2482}}
& \multicolumn{1}{c}{\multirow{2}{*}{2267}}  &\multicolumn{1}{c}{\multirow{2}{*}{2522.3}} &\multicolumn{1}{c}{--}\\
\rho_3(2^3G_3)   &    &\multicolumn{1}{c}{ }   & \multicolumn{1}{c}{}&\multicolumn{1}{c}{ }&\multicolumn{1}{c}{--} \\
\omega_3(3^3G_3) &    &\multicolumn{1}{c}{\multirow{2}{*}{2746}}
& \multicolumn{1}{c}{\multirow{2}{*}{ }}  &\multicolumn{1}{c}{\multirow{2}{*}{2748.5}} &\multicolumn{1}{c}{--}\\
\rho_3(3^3G_3)   &    &\multicolumn{1}{c}{ }   & \multicolumn{1}{c}{}&\multicolumn{1}{c}{ }&\multicolumn{1}{c}{--} \\
\hline
\hline
\end{array}\]
\end{center}
\end{table*}

\subsection{Decay modes and width}\label{sec decay}

 $\omega$ and $\omega_3$ and $\rho$ and $\rho_3$ mesons have the same quantum numbers ($P$ and $C$), for which reason the study of them can be learned from each other. We estimate the {\color{black}masses} of the missing states in these meson families. Similarly, we study the strong {\color{black}decays} of $\omega$ and $\omega_3$ meson family comparing with that of $\rho$ and $\rho_3$ meson family. The $\gamma$ value in Eq. (\ref{gamma}) is taken by the following method.  When fitting the $\rho$ and $\rho_3$ meson experimental width value (with error) using the theoretical total width, the range of $\gamma$ value can be fixed. Then we use this $\gamma$ value to calculate the width of  $\omega$ and $\omega_3$ meson. Some decay channels which less than 1 MeV are omitted.
\subsubsection{\texorpdfstring{$S$}~wave~\texorpdfstring{$\omega$}~~and~ \texorpdfstring{$\rho$}~~mesons}

\begin{table}[htbp]
\centering
\caption{The total and partial decay widths of the  $\omega(2S)$ and $\rho(2S)$, the unit of width is MeV. The $\gamma$ value is 11.5$-$13.4. \label{decay2s}}
\[\begin{array}{ccccccc}
\hline
\hline
\multicolumn{3}{c}{\omega(1420),~ {\Gamma_{exp.}=880\pm170}$~\cite{Aulchenko:2015mwt}$}&\multicolumn{3}{c}{{\rho(1450)},~ {\Gamma_{exp.}=400\pm60}$~\cite{Tanabashi:2018oca}$}\\
\hline
\text{Channel} &\text{Value}  &\text{Ref.~\cite{Barnes:1996ff}} &\text{Channel}  &\text{Value}  &\text{Ref.~\cite{Barnes:1996ff}} \\\midrule[1pt]
\text{Total}    &690\pm105    &{378}        &\text{Total}       &400       &{279}\\
\pi\rho  &{608}        &{328}        &\pi\omega   &{220}   &{122}\\
\eta\omega &{33.3}       &{   }        &\eta\rho    &{58.0}   &{   }\\
KK       &{28.3}       &{31}         &KK          &{31.9}   &{   }\\
{b_1}\pi &{12.1}       &{1}          &\pi\pi      &{30.9}   &{74}\\
KK^*     &{9.50}       &{5}          &KK^*        &{30.3}   &{   }\\
{  }     &             &{   }        &{a_1}\pi    &{14.9}   &{3  }\\
{  }     &             &{   }        &{h_1}\pi    &{9.05}    &{1  }\\
 \hline
 \hline
\end{array}\]
\end{table}
By analyzing the above masses of $\omega$ and $\rho$ states, we know that  $\omega(1420)$,  $\omega(1960)$, $\omega(2330)$, $\omega(2290)$, and $X(2240)$ are the radial excitations of $\omega(782)$.
$\rho(1450)$, $\rho(1900)$, and $\rho(2150)$ are the $\rho(2S)$, $\rho(3S)$, and $\rho(4S)$ states respectively. We will then discuss these states.

\par
 From Table \ref{decay2s}, $\omega(1420)$ dominantly decay into {\color{black}$\pi\rho$} which is consistent with experiment \cite{Tanabashi:2018oca} and Ref. \cite{Barnes:1996ff}. {\color{black}$\eta\omega$},  $KK$,  ${b_1}\pi$ and $KK^*$ are the important decay modes in which ${b_1}\pi$ was observed in experiment \cite{Tanabashi:2018oca}.
 The total width of $\omega(1420)$ in our calculation has a overlap with the experiment value \cite{Aulchenko:2015mwt}. $\omega(1420)$ is a good candidate of $\omega(2S)$ {\color{black}which is consistent} with Ref. \cite{Wang:2012wa}.
 \par

\begin{table}[H]
\centering%
\caption{The total and partial decay width of the  $\omega(3S)$  and $\rho(3S)$, the unit of  width is MeV. The $\gamma$ value is 6.3$-$7.9.   \label{decay3s}}
\vspace{-0.2cm}
\[\begin{array}{cccc}
\hline
\hline
\multicolumn{2}{c}{\omega(1960),~{\Gamma_{exp.}=195\pm60}$~\cite{Anisovich:2011sva}$}&\multicolumn{2}{c}{\rho(1900),~ {\Gamma_{exp.}=130\pm30}$~\cite{Tanabashi:2018oca}$}\\
\hline
\text{Channel}&\text{Value}   &\text{Channel}  &\text{Value} \\ \midrule[1pt]
\text{Total}  &{191\pm42.5}   &\text{Total} &130\\
\rho\pi       &{108}         &{a_2}\pi     &{31.1}\\
{b_1}\pi      &{32.0}        &{\color{black}\pi\omega}    &{29.2}\\
K{K_1}        &{14.8}        &K{K_1}       &{18.7}\\
\eta\omega    &{9.90}        &\pi\pi(1300) &{16.1}\\
KK^*(1410)    &{9.40}        &\pi\pi       &{13.9}\\
{h_1}\eta     &{4.74}        &{a_1}\pi     &{13.0}\\
 KK^*         &{3.77}        &KK           &{2.60}\\
KK            &{3.14}        &{b_1}\eta    &{2.55}\\
K{K_1^\prime} &{2.35}        &KK^*         &{2.46}\\
 \hline
 \hline
\end{array}\]
\end{table}

\begin{table*}[hbp]
\centering%
\caption{The total and partial decay widths of the  $\omega(4S)$ and $\rho(4S)$,  the unit of width is MeV. The $\gamma$ value is 6.9$-$11.3.           \label{decay3}}
\[\begin{array}{cccccc}
\hline
\hline

                 &{\omega(2330)}  &{\omega(2290)}    &{X(2240)}  &\multicolumn{2}{c}{\rho(2150)}   \\
\hline
\text{Channel}&\text{Value}&\text{Value}&\text{Value}&\text{Channel} & \text{Value}   \\ \midrule[1pt]
\Gamma_{exp.}     &$435$ \pm $75~\cite{Zyla:2020zbs}$  &$275$ \pm $35~\cite{Zyla:2020zbs}$ &$139.8$ \pm $23.99~ \cite{Ablikim:2018iyx}$ &\Gamma_{exp.}  &$310$ \pm $140~\cite{Zyla:2020zbs}$\\
\text{Total}      &{372\pm170}     &346\pm158  &306\pm140 &\text{Total} &{310}\\
{\color{black}\pi\rho(1450)}     &{134}   &122    &95.8   &\pi\omega_3     &{58.6}    \\
{\color{black}\pi\rho}           &{62.6}  &68.0   &69.0   &{a_2}\pi(1700)  &{40.7}   \\
{b_1}\pi          &{61.6}  &54.4   &42.4   &\pi\pi(1300)    &{32.1}  \\
{a_1}\rho         &{37.2}  &39.0   &37.3   &\pi\omega       &{31.5}   \\
{\color{black}\pi\rho_3}         &{19.5}  &17.3   &12.9   &a_2\pi          &{23.1}   \\
K{K_1}            &{8.39}  &8.33   &6.69   &h_1\pi          &{14.7}   \\
f_1\omega         &{7.84}  &6.28   &4.28   &a_1\pi          &{13.6}   \\
{\color{black}{h_1}\eta}         &{7.01}  &7.53   &7.54   &b_1\rho         &{11.5}    \\
K^*K^*            &{5.91}  &5.65   &4.88   &\pi\pi          &{11.4}     \\
\eta\omega        &{4.81}  &3.28   &1.70   &a_1\omega       &{10.5}     \\
KK(1460)          &{4.81}  &3.45   &2.00   &{\color{black}\pi\omega}       &{8.31}     \\
{a_2}\rho         &{2.65}  &11.5   &19.0   &\rho\rho        &{7.74}     \\
KK                &{0.599} &0.503  &0.379  &{\color{black}{K}K_1^\prime}  &{6.47}      \\
KK_1^\prime       &{0.421} &0.224  &0.0783  &f_2\rho         &{6.19}     \\
K^*{K_1}          &0.164   &0.152  &0.0761  &{\color{black}\pi\pi_2}        &{5.98}     \\
KK^*(1430)        &0.121   &0.0366  &0.00138&{\color{black}{a_0}\omega}     &{5.49}     \\
{f_2}\omega       &{0.0228} &1.39   &4.97   &a_2\omega       &{5.34}      \\
                  &         &        &        &f_1\rho         &{4.99}      \\
                  &         &        &        &K^*K^*          &{4.20}      \\
                  &         &        &        &{\color{black}KK^*(1410)}      &{2.92}     \\
                  &         &        &        &b_1\eta         &{1.74}   \\
                  &         &        &        &KK(1460)        &{1.20}     \\
                  &         &        &        &{\color{black}\eta\rho}        &{0.99}    \\
                  &         &        &        &\eta\rho(1450)  &0.903\\
 \hline
 \hline
\end{array}\]
\end{table*}

$\omega(1960)$ is observed in the $p\bar{p}\to \omega\eta$, and $\omega\pi\pi$ process \cite{Anisovich:2011sva}.
As shown in Table \ref{decay3s}, $\omega(1960)$ dominantly decay into $\rho\pi$ under $\omega(3S)$ assignment. The decay modes ${b_1}\pi$, $KK_1$ and {\color{black}$\eta\omega$ make some contribution} to the total width, and ${b_1}\pi$ can decay to $\omega\pi\pi$ which is the final channel observed in experiment \cite{Anisovich:2011sva}. According to our calculation, the total width of $\omega(1960)$ is $191.1 \pm 42.5$ MeV and this is consistent with the experimental data \cite{Anisovich:2011sva}.  In addition, {\color{black}$\eta\omega$} is a sizable final channel and has been observed in experiment \cite{Anisovich:2011sva}. Other detailed information is demonstrated in Table \ref{decay3s}. Our calculation indicates that $\omega(1960)$ can be assigned as $\omega(3S)$ state, {\color{black}which is consistent with Ref. \cite{feng:2021igh}.}

$X(2240)$ is observed in the $e^+e^- \to K^+K^-$ process by the BESIII Collaboration, which has the mass of $2239.2 \pm 7.1 \pm 11.3 \, \rm{MeV}$ and the width of 139.8$\pm$12.3$\pm$20.6 MeV \cite{Ablikim:2018iyx}.
The quantum number of this resonant structure can be assigned as $J^{PC} = 1^{--}$. Assuming that it is a isospin scalar state, it may be a $\omega(4S)$ candidate from our previous mass analysis. The state $X(2240)$ needs more theoretical and experimental research to recover  its structure.
$\omega(2330)$ is firstly observed in the $\gamma{p} \to \rho^\pm\rho^0\pi^\mp$ process \cite{valenciarho}. It has the mass of $2330 \pm 30$ MeV and the width of $435 \pm 75$ MeV.
$\omega(2290)$ is reported in the partial wave analysis of the data on $p\bar{p} \to \bar{\Lambda}\Lambda$ \cite{Bugg:2004rj}, which has the mass of $2290 \pm 20$ MeV and the width of $275 \pm 35$ MeV.
$\omega(2330)$, $\omega(2290)$ and $X(2240)$ are the candidates of the $\omega(4S)$ according to our calculation and the obtained widths are in good agreement with experimental values. {\color{black}$\pi\rho(1450)$} is the primary decay mode of $\omega(2330)$, $X(2240)$ and $\omega(2290)$ if assign them as $\omega(4S)$ states. Besides, {\color{black}$\pi\rho$, $b_1\pi$ and $a_1\rho$} make needful contribution to the total width of $\omega(4S)$, other decay modes contribute less to the total width as shown in Table \ref{decay3}.

As for $S$ wave {\color{black}states} of $\rho$ family, we assign $\rho(1450)$, $\rho(1900)$ and $\rho(2150)$ as $\rho(2S)$, $\rho(3S)$ and $\rho(4S)$ states, which consists with the results of {\color{black}Ref. \cite{Ebert:2009ub, Afonin:2007aa, Wang:2021abg, He:2013ttg}.}
When $\rho(1450)$ is treated as the $\rho(2S)$, {\color{black}$\pi\omega$} and $\eta\rho$ are predicted to be its important decay channels. $KK$, $\pi\pi$ and $KK^*$ are the sizable decay modes as well. $\rho(1900)$ is a good candidate of $\rho(3S)$. As shown in the second column of Table \ref{decay3s}, the decay modes of {\color{black}$\rho(1900)$} are predicted. {\color{black}$a_2\pi$, $\pi\omega$ and $KK_1$} are its important decay modes. {\color{black}$\pi\pi(1300)$, $\pi\pi$} and $a_1\pi$ are the sizable final states, {\color{black}which is consistent with Ref. \cite{Wang:2020kte}}. We assign $\rho(2150)$ as $\rho(4S)$ based on our calculation. The channels $\pi\omega_3$, $a_2(1700)\pi$, $\pi\pi(1300)$ and $\pi\omega$ are the main channels. Otherwise, {\color{black}$a_2\pi$, $h_1\pi$,  $a_1\pi$ and ${b_1}\rho$} have sizable contributions to the total width of $\rho(4S)$.

\subsubsection{D-wave \texorpdfstring{$\omega$}~~and~ \texorpdfstring{$\rho$}~~mesons}
In this section, we will give an analysis for the D-wave $\omega$ and $\rho$ mesons
by the means of their two-body strong decay {\color{black}behaviors.}
\begin{table}[ht]
\centering%
\caption{The partial decay widths of the $\omega(1D)$, $\rho(1D)$,  the unit of  width is MeV. The $\gamma$ value is 4.14$-$6.3.      \label{decay4}}
\[\begin{array}{cccccc}
\hline
\hline
\multicolumn{3}{c}{\omega (1650),~ {\Gamma_{exp.}=315\pm35}$~\cite{Tanabashi:2018oca}$}&\multicolumn{3}{c}{\rho(1700),~{\Gamma_{exp.}=250\pm100}$\cite{Tanabashi:2018oca}$}\\
\hline
\text{Channel} &\text{Value} &\text{Ref.~\cite{Barnes:1996ff}}& \text{Channel}  &\text{Value} &\text{\cite{Barnes:1996ff}}\\ \midrule[1pt]
\text{Total }&284\pm113 &{542}  &\text{Total} &{250}    &{435}\\
{b_1}\pi     &{272}     &{371}  &{a_1}\pi     &{93.9}  &{134}\\
{\color{black}\pi\rho}      &{42.1}    &{101}  &{h_1}\pi     &{85.2}  &{124}\\
KK           &{9.00}    &{35}   &\rho\rho     &{26.4}  &{14}\\
\eta\omega   &{7.84}    &{ }    &{\color{black}\pi\omega}    &{14.1}  &{35} \\
KK^*         &{5.29}    &{21}   &KK           &{9.84}  &{36}\\
{}           &{   }     &{ }    &\eta\rho     &{8.76}  &{}\\
{}           &{   }     &{ }    &KK^*         &{6.69}  &{}\\
{}           &{   }     &{ }    &{a_2}\pi     &{2.64}  &{}\\
 \hline
 \hline
\end{array}\]
\end{table}

 $\omega(1650)$ is well established as a $1D$ state in the $\omega$ family in theory \cite{Wang:2012wa,Tanabashi:2018oca,Wang:2019jch}, and its dominant decay channel is $b_1\pi$, whose branching ratio is about 0.8. {\color{black}$\pi\rho$}, $KK$, $\eta\omega$ and $KK^*$ are the important modes.
 $\rho(1700)$ is a good candidate of the $1^3D_1$ $\rho$ meson. The analysis of $\rho(1700) \to 2\pi, 4\pi$ \cite{20014} and the study of {\color{black}$e^+e^- \to \pi\omega$} via the nonrelativistic $^3P_0$ quark model \cite{2009E} both indicate that $\rho(1700)$ is a $1^3D_1$ state. $\rho(1700)$ mainly decays into $a_1\pi$, $h_1\pi$ and $\rho\rho$  with branching ratios being about 0.38, 0.34 and 0.11, respectively.

\begin{table}[H]
\centering%
\caption{The total and partial decay widths of the  $\omega(2D)$ and $\rho(2D)$,  the unit of  width is MeV. The $\gamma$ value is 5.4$-$6.4.    \label{decay2D}}
\[\begin{array}{cccc}
\hline
\hline
\multicolumn{2}{c}{\omega(2D)}
&\multicolumn{2}{c}{\rho(2000),~{\Gamma=260\pm45}$~\cite{Tanabashi:2018oca}$}\\
\hline
 \text{Channel}     &\text{ Value}   &  \text{Channel}& \text{ Value} \\ \midrule[1pt]
 \text{Total}       &{212\pm36.1}& \text{Total}     &{260} \\
{b_1}\pi            &{110}         &{a_1(1640)}\pi  &58.4\\
\pi\rho(1450)       &{24.7}         &{\color{black}{h_1}\pi}       &{36.6}\\
\pi\rho             &{22.7}         &\rho\rho       &{35.2}\\
{\color{black}{a_1}\rho}           &{22.1}         &\pi\pi_2       &{34.1}\\
{\color{black}{h_1}\eta}           &{13.2}         &{\color{black}{a_1}\pi}       &{26.3}\\
KK^*(1410)          &{4.29}         &\pi\pi            &{21.9}\\
\eta\omega(1420)    &{3.26}         &\pi\pi(1300)      &{19.8}\\
\eta\omega          &{2.60}         &\pi\omega(1420)   &{8.22}\\
KK                  &{2.18}         &\pi\omega         &{6.39}\\
KK_1                &1.83           &{\color{black}{b_1}\eta}         &{4.50}\\
KK^*                &1.15           &KK^*(1410)        &2.54\\
                     &                 &KK                &{1.81}\\
                     &                &\eta\rho          &1.70\\
 \hline
 \hline
\end{array}\]
\end{table}

As is shown in Fig. \ref{regge} and Table \ref{mass}, the mass of $\omega(2D)$ is predicted, which is still unobserved. $\omega(2D)$ has the total width of about {\color{black}$212 \pm 36.1$ MeV.} $b_1\pi$ is its dominant decay channel, whose branching ratio is 0.5. $\pi\rho(1450)$, $\pi\rho$ and {\color{black}${a_1}\rho$} are its important decay channels and more details can be seen in Table \ref{decay2D}. We suggest that experimentalists focus on these final channels.
In 1994, a resonance around 1988 MeV was found in $\bar{p}p \to \pi\pi$ from 0.36 to 2.5 GeV reaction {\color{black}\cite{Hasan:1994he}}. Later in 2011, Anisovich $et~al$. found a resonance that $J^{PC} = 1^{--}$ and its mass is of $2000 \pm 30$ MeV {\color{black} \cite{Anisovich:2000ut}}. $\rho(2000)$ is treated as $\rho(2D)$ state {\color{black} \cite{He:2013ttg,Wang:2020kte}}. $a_1(1640)\pi$, {\color{black}${h_1}\pi$}, $\rho\rho$, and $\pi\pi_2(1670)$ have sizable contributions to the total width of $\rho(2D)$.

\begin{table}[hbtp]
\centering%
\caption{The partial decay widths of the  $\omega(3D)$, $\rho(3D)$,  $\Gamma_exp.$ means the experimental total width, the unit of width is MeV.  The $\gamma$ value is 11.3$-$14.6.  \label{decay3D}}
\[\begin{array}{ccccc}
\hline
\hline
        &\omega (2205)     &X(2220)            &\multicolumn{2}{c}{\rho(2270)}   \\ \hline
 \text{Channel}      &\text{Value}      &\text{ Value}      &\text{Channel }  &\text{Value}\\
 \midrule[1pt]
\Gamma_{exp.}        &$350$ \pm $90~\cite{Zyla:2020zbs}$          &$59$ \pm $30.6~ \cite{BESIII:2021uni}$          &\Gamma_{exp.}    &$325$ \pm $80~\cite{Anisovich:2011sva}$   \\
\text{Total}         &147\pm36.1     &164\pm40.4      &\text{Total}     &325   \\
{b_1}\pi             &{56.7}           &66.0              &{\color{black}{a_2}\omega}      &67.0\\
{\pi}\rho(1450)      &{21.0}           &23.7              &\pi\pi(1300)      &64.3\\
{\color{black}{a_1}\rho}            &{23.7}           &22.0        &{\color{black}{h_1}\pi}         &42.2\\
{\color{black}{a_2}\rho}           &{9.08}           &12.2              &\pi\pi           &39.4 \\
{\color{black}\pi\rho}              &{8.32}           &9.78       &{\color{black}{a_1}\pi}         &26.1\\
\pi\rho_3            &{6.09}           &6.97              &\pi\pi_2         &23.2\\
{\color{black}{f_2}\omega}          &{5.20}           &6.59              &\rho\rho         &19.5\\
{\color{black}{f_1}\omega}          &{5.24}           &5.54              &{\color{black}{f_2}\rho}       &12.0\\
KK_1                 &{4.78}           &4.58              &{a_1}\omega      &5.81\\
\pi\rho(1700)        &{2.65}           &2.90              &{\color{black}{f_1}\rho}        &5.67\\
{\color{black}{h_1}\eta}            &{2.49}           &3.41              &\omega\pi        &4.51\\
{\color{black}\pi(1300)\rho}        &{1.34}           &0.495              &KK_1       &4.00\\
                    &                   &                   &\rho\rho(1450)   &4.00\\
                     &                   &                  &\pi\omega_3&{3.41}\\
                     &                   &                  &{\color{black}{b_1}\eta}  &2.58 \\
                     &                   &                  &{\color{black}\pi\omega(1650)}  &{1.5}\\
\hline
 \hline
\end{array}\]
\end{table}

$X(2220)$ state is found in the $e ^+e^- \to \omega\pi^0\pi^0$ process \cite{BESIII:2021uni}, which  has the mass of $M = 2222\pm7\pm2$ MeV and the width of $\Gamma = 59\pm30\pm6$ MeV.
$X(2220)$ is an excellent candidate of $\omega(3D)$ by Regge trajectory analysis and according to Table \ref{decay3D}.
When $X(2220)$ is treated as the $\omega(3D)$ state, its width is close to the experiment value.
$\omega(2205)$ is also assigned as $\omega(3D)$ state based on mass {\color{black}spectra} analysis.
$b_1\pi$ is the main decay mode of $\omega(3D)$. $\pi\rho(1450)$ and {\color{black}${a_1}\rho$} are the important decay channels of $\omega(3D)$, other modes like {\color{black}${a_2}\rho$, $\pi\rho$ and $\pi\rho_3$} and so on also have contribution to the total width of $\omega(3D)$.
{\color{black}We assign $\rho(2270)$ as $\rho(3D)$ state, which is consistent with Ref. \cite{He:2013ttg}. $\rho(2270)$ primarily decays into ${a_2}\omega$, $\pi\pi(1300)$ and ${h_1}\pi$.} More details can be seen in Table \ref{decay3D}.

\subsubsection{D-wave~ \texorpdfstring{$\omega_3$}~~and~~\texorpdfstring{$\rho_3$}~~mesons}
{\color{black}Referring to Particle Data Group (PDG)}, we can find some experimental results about $\omega_3$ and $\rho_3$, which are  $\rho_3(1690)$, $\rho_3(1990)$, $\rho_3(2250)$, $\omega_3(1670)$, $\omega_3(1945)$, $\omega_3(2285)$, and $\omega_3(2255)$. We carry out an investigation of the Okubo-Zweig-Iizuka allowed two-body strong decay of the {\color{black}$D$-wave and $G$-wave} $\omega_3$ and $\rho_3$ mesons.

\begin{table}[ht]
\centering%
\caption{The partial decay widths of the $\omega_3(1D)$, $\rho_3(1D)$, the unit of width is MeV. The $\gamma$ value is 7.5$-$9.2.      \label{decay1DD}}
\[\begin{array}{cccc}
\hline
\hline
\multicolumn{2}{c}{\omega_3(1670),~ {\Gamma_{exp.}=90 \pm 20}~$\cite{1969Evidence}$}  &\multicolumn{2}{c}{\rho_3(1690),~{\Gamma_{exp.}= 190 \pm 40}~$\cite{Zyla:2020zbs}$}\\
\hline
\text{Channel}&\text{Value}&\text{Channel}&\text{Value} \\\midrule[1pt]
\text{Total } &{81.8\pm16.5}    &\text{Total}  &{190}    \\
\pi\rho       &{62.8}            &\rho\rho      &{105}     \\
b_1\pi        &{14.7}             &\pi\pi        &{32.0}    \\
\eta\omega    &{2.76}             &\pi\omega     &{23.1}    \\
KK            &{1.16}             &\pi{h_1}&{10.6}    \\
              &                  &{\color{black}{a_2}\pi} &{8.07}     \\
              &                    &\eta\rho      &{3.68}     \\
              &{ }                 &{\color{black}{a_1}\pi}&{3.26}     \\
              &{ }                 &KK            &{2.66}     \\
 \hline
 \hline
\end{array}\]
\end{table}
There exists a lot of data for $\omega_3(1670)$ from PDG. Amelin $et~al$. first found it in the $\pi^-p \to \pi^+\pi^-\pi^0n$ reactions, which has the mass of $1665.3 \pm 5.2 \pm 4.5$ MeV \cite{1996Partial}.
{\color{black}And} then {\color{black}Baltay $et~al$.} reported the existence of a high-mass, $I=0$ resonant state with odd $G$ parity, whose mass and width are $M = 1.63 \pm 0.012$ GeV and $\Gamma = 0.173 \pm 0.016$ GeV {\color{black} \cite{1978Production}}, respectively.  At present, $\omega_3(1670)$ is {\color{black}well} established to be a $1^3D_3$ state \cite{Buisseret:2004wm}. $\pi\rho$ is the dominant decay modes of $\omega_3(1D)$, which has the branching ratios of 0.77. Additionally, {\color{black}$b_1\pi$, $\eta\omega$, and $KK$} also make contribution to the total width of $\omega_3(1D)$.
{\color{black} $\rho_3(1690)$ is well established to be a $1^3D_3$ state, which decays into $\rho\rho$, $\pi\pi$, $\pi\omega$, and $\pi{h_1}$ as shown in Table \ref{decay1DD} and the branching ratios are about 0.55, 0.17, 0.12, and 0.056, respectively. These results are in good agreement with that in Ref. \cite{He:2013ttg}. }

$\omega_3(1945)$ was firstly observed in the $p\bar{p} \to \omega\eta, \omega\pi^0\pi^0$ \cite{Anisovich:2002xoo} .
It has the mass of $1945 \pm 20$ MeV and the width of $115 \pm 22$ MeV. We assign it as $\omega_3(2D)$ state and calculated the two-body strong decay as shown in Table \ref{decay2DD}. We see from Table \ref{decay2DD} that $b_1\pi$, $\pi\rho$, $\pi\rho(1450)$ and {\color{black}$\pi\rho_3$} are major decay modes of $\omega_3(2D)$. $\eta\omega$, $h_1\eta$, $K^*K^*$ and $KK^*$ are important channels. When  treated as $\rho_3(2D)$, $\rho_3(1990)$ mainly decay into $\pi\pi(1300)$, {\color{black}${a_2}\pi$}, $\rho\rho$, which branching ratios are 0.22, 0.18 and 0.14, respectively {\color{black}and this result was certified in Ref. \cite{He:2013ttg}. Besides, $\rho_3(1990) \to \pi{a_1}, \pi{h_1}, \pi\omega(1420)$} are sizeable. More details can be seen in Table \ref{decay1DD}.

As for $\omega_3(3D)$, we consider $\omega_3(2285)$ is the best candidate according to the mass spectra analysis and we give its decay modes and values in Table \ref{decay3DD}. The total width of $\omega_3(2285)$ is $226 \pm 126$ MeV if we treat $\omega_3(2285)$ as a $\omega_3(3D)$ state, {\color{black}which is in great agreement with experimental width \cite{Bugg:2004rj}}. $a_2\rho$, $\pi\rho$, $f_2\omega$ and $b_1\pi$ make great contribution to its total width and {\color{black}$\pi\rho_3$}, $\pi(1300)\rho$, $\pi\rho(1450)$ and so forth are also important decay channels. Meanwhile, $\rho_3(2250)$ is established to be a $\rho_3(3D)$ state, which can decay into {\color{black}${a_2}\omega$, ${f_2}\rho$,} $\pi\omega$, $\rho\rho$ as shown in Table \ref{decay3DD}.

$\omega_3(4D)$ and $\rho_3(4D)$ are still missing in experiment. We calculated the width of $\omega_3(4D)$ is about 290 MeV and $\rho_3(4D)$ is approximately 420 MeV. According to Table \ref{decay34D}, the main decay modes of $\omega_3(4D)$ are $a_2\rho$, $\pi(1300)\rho$, $b_1\pi$ and $\pi\rho$. $\rho_3(4D)$ mainly decay into {\color{black}${a_4(2040)\pi}$, ${a_2(1700)\omega}$, ${a_2(1700)\pi}$} and so on. These predictions can help us search for $\omega_3(4D)$ and $\rho_3(4D)$ states.

\begin{table}[ht]
\centering%
\caption{The partial decay widths of the $\omega_3(2D)$, $\rho_3(2D)$,  the unit of width is MeV. The $\gamma$ value is 10.4$-$11.8.      \label{decay2DD}}
\[\begin{array}{cccc}
\hline
\hline
\multicolumn{2}{c}{\omega_3 (1945),~ {\Gamma_{exp.}=115 \pm 22}~$\cite{Zyla:2020zbs}$}  &\multicolumn{2}{c}{\rho_3(1990),~{\Gamma_{exp.}=188 \pm 24}~$\cite{Anisovich:2011sva}$}\\\hline
\text{Channel}  &\text{Value } &\text{Channel}&\text{Value}     \\ \midrule[1pt]
\text{Total }   &{105\pm13.2}        &\text{Total}     &{188}\\
{b_1}\pi        &{38.5}               &\pi\pi(1300)     &{40.6}\\
\pi\rho         &{22.1}               &{\color{black}{a_2}\pi}   &{34.2}\\
\pi\rho(1450)   &{18.7}               &\rho\rho         &{26.9}\\
{\color{black}\pi\rho_3}      &{11.5}               &{\color{black}{a_1}\pi}   &{16.7}\\
\eta\omega      &{4.51}               &{\color{black}{h_1}\pi}   &{16.4}\\
h_1\eta         &{4.44}               &\pi\omega(1420)  &{15.5}\\
K^*K^*          &{2.31}               &\pi\pi_2   &{9.19}\\
KK^*            &{1.28}               &\pi\omega_3&{7.80}\\
                &{ }                   &\pi\omega        &{7.14}\\
                &{ }                   &\eta\rho         &{4.58}\\
                &{ }                   &{\color{black}{b_1}\eta}        &{2.63}\\
                &{ }                   &K^*K^*           &{2.54}\\
                &{ }                   &KK^*             &{1.61}\\
 \hline
 \hline
\end{array}\]
\end{table}

\begin{table}[ht]
\centering%
\caption{The partial decay widths of the $\omega_3(3D)$, $\rho_3(3D)$,  the unit of width is MeV. The $\gamma$ value is 6.7$-$12.5.      \label{decay3DD}}
\[\begin{array}{cc|cc}
\hline
\hline
\multicolumn{2}{c}{\omega_3 (2285),~ {\Gamma_{exp.}=224 \pm 50}~$\cite{Bugg:2004rj}$}  &\multicolumn{2}{c}{\rho_3(2250),~{\Gamma_{exp.}=135 \pm 75}~$\cite{1977Antiproton}$}\\
\hline
\text{Channel}  & \text{Value }     & \text{Channel}    &\text{Value}\\ \midrule[1pt]
\text{Total }       &{226\pm126}   &\text{Total}        &{135}     \\
a_2\rho             &{76.3}           &{\color{black}{a_2}\omega}      &{58.5}  \\
\pi\rho             &{32.6}           &{\color{black}{f_2}\rho}      &{33.7}  \\
f_2\omega           &{25.6}           &\pi\omega      &{12.3}  \\
b_1\pi              &{24.5}           &\rho\rho       &{11.6}  \\
{\color{black}\pi\rho_3}           &{12.9}           &{\color{black}{h_1}\pi}       &{11.0}  \\
\pi(1300)\rho       &{10.8}           &{\color{black}{a_2}\pi}       &{9.26}  \\
\pi\rho(1450)       &{5.22}           &{\color{black}{a_1}\pi}       &{8.60}  \\
a_1\rho             &{4.37}           &\pi\omega_3    &{4.32}  \\
\eta\omega          &{3.78}           &\pi\pi         &{3.68}  \\
\eta\omega(1420)    &{3.68}           &\pi\pi_2       &{2.95}  \\
h_1\eta             &{3.28}           &\pi\omega(1420)&{2.70}  \\
\eta(1295)\omega    &2.14             &{\color{black}{b_1}\eta}      &{2.09}  \\
K^*K^*              &{1.45}           &KK^*(1410)     &{2.06}  \\
                    &                  &{\color{black}{a_1}\omega}      &{1.92}  \\
                    &                  &\eta\rho       &{1.82}  \\
                    &                  &{\color{black}\eta(1295)\rho} &{1.81}  \\
                    &                  &\eta\rho(1450) &{1.70} \\
                    &                  &K^*K^*         &{1.69}  \\
 \hline
 \hline
\end{array}\]
\end{table}

\begin{table}[ht]
\centering%
\caption{The partial decay widths of the $\omega_3(4D)$, $\rho_3(4D)$,  the unit of width is MeV. The $\gamma$ value is 11.6.      \label{decay34D}}

\[\begin{array}{cccc}
\hline
\hline
\multicolumn{2}{c}{\omega_3(2630)}&\multicolumn{2}{c}{\rho_3(2630)}\\
\hline
\text{Channel}      &\text{Value } &\text{Channel}         &\text{Value}  \\ \midrule[1pt]
\text{Total }       &{288}       &\text{Total}           &{420}        \\
a_2\rho             &57.2         &a_4(2040)\pi           &184\\
 \pi(1300)\rho      &40.3         &a_2(1700)\omega        &38.7\\
 b_1\pi             &37.3         &a_2(1700)\pi           &35.4\\
 \pi\rho            &33.3         &a_2\pi                 &25.3\\
 \pi\rho(1450)      &32.8         &\rho\rho(1450)         &22.4\\
 {\color{black}\pi\rho_3}          &23.0         &f_2\rho                &16.9\\
 f_2\omega          &17.2         &h_1\pi                 &14.3\\
 \eta(1295)\omega   &8.03         &b_1\rho                &11.8\\
\pi_2\rho           &6.85         &{\color{black}\pi\omega(1420)}        &10.2\\
\eta\omega(1420)    &6.42         &{\color{black}\pi\omega_3}            &8.75\\
 K^*K^*(1410)       &5.65         &\rho\rho               &6.76\\
h_1\eta             &5.30         &{\color{black}\pi\pi_2}               &6.65\\
\eta\omega          &3.46         &K^*K^*(1410)           &5.65\\
K^*K_2^*            &2.36         &{\color{black}\eta\rho(1450)}         &5.51\\
f_2h_1              &1.94         &\pi\pi                 &4.79\\
\eta_2\omega        &1.86         &b_1\eta                &4.01\\
f_1\omega           &1.66         &a_1\omega              &3.27\\
a_2{b_1}            &1.50         &{\color{black}K^*K_2^*}               &2.36\\
f_1h_1              &1.09         &\pi_2\omega            &2.14\\
\pi\rho(1700)       &1.01         &h_1{a_1}               &1.94\\
                    &              &b_1{b_1}               &1.83\\
                    &              &{\color{black}KK^*(1410)}             &1.71\\
                    &              &a_2h_1                 &1.51\\
                    &              &{\color{black}\eta^\prime\rho(1450)}  &1.39\\
                    &               &\pi\pi(1300)           &1.21\\
                    &               &h_1\pi(1300)           &1.04\\
 \hline
 \hline
\end{array}\]
\end{table}

\subsubsection{G-wave \texorpdfstring{$\rho_3$}~~and~~ \texorpdfstring{$\omega_3$}~~mesons}
{\color{black}$\omega_3(2285)$ is regarded as $\omega_3(1G)$ in Ref. \cite{Afonin:2007aa}. The reference \cite{Anisovich:2011sva} observed $\omega_3(2255)$ and assigned it as $\omega_3(1G)$ state. According to the mass spectra analysis and two-body decay modes,  we agree that $\omega_3(2255)$ may be $\omega_3(1G)$ state.} According to {\color{black}Table} \ref{decay31G}, $b_1\pi$, {\color{black}$\pi\rho_3$} and $h_1\eta$ are predicted to be $\omega_3(1G)$'s important decay channels, which have the ratios of 0.45, 0.14 and 0.1, respectively. We suggest that experimentalists focus on these final channels. The total width of  $\omega_3(1G)$ is 325 MeV. The total width of $\rho_3(1G)$ is approximately 721 MeV. {\color{black}$\pi\pi_2$}, $b_1\rho$, $\rho\rho$ and $a_1\pi$ are its important decay channels. Considering the final decay channels, {\color{black}$\pi\pi_2$} will be the most important final channels in searching for $\rho_3(1G)$ state experimentally. More details can be seen in Tale \ref{decay31G}. These predictions can help us search for and establish $\omega_3(1G)$ and $\rho_3(1G)$ state.

{\color{black}$\omega_3(2G)$ and $\rho_3(2G)$ and $\omega_3(3G)$ and $\rho_3(3G)$ are still missing, We make predictions about their widths and decay modes.}
The decay information of the $\omega_3(2G)$ and $\rho_3(2G)$ is listed in Table \ref{decay32G}. As shown in the first column of Table \ref{decay32G}, the strong decay of $\omega_3(2G)$ is predicted, which is still unobserved.
$\omega_3(2G)$ has the total width of 312 MeV. $b_1\pi$ is its dominant decay channel, which ratios is 0.34. $a_2\rho$, $\pi_2\rho$, {\color{black}$\pi\rho_3$}, $h_1\eta$ and $\pi\rho$ are the important final states. {\color{black}$\eta^\prime\omega$}, $a_1\rho$ and $h_1\eta^\prime$ are small. The total width of $\rho_3(2G)$ is 397 MeV. {\color{black}$\rho_3(2G) \to \pi\pi_2$} will be the dominant decay mode. In the calculation, $h_1\pi$, $a_1\pi$ and $\rho\rho(1450)$ are the important decay channels. Other decay information can be seen in Table \ref{decay32G}.

The decay information of the $\omega_3(3G)$ and $\rho_3(3G)$ are also predicted in this work {\color{black}in Table \ref{decay33G}}. The total width of $\omega_3(3G)$ is approximately 181 MeV. The channels $b_1\pi$, $\pi\rho$, {\color{black}$\pi\rho_3$} and $f_1\omega$ have the branching ratios of 0.35, 0.07, 0.07 and 0.06, respectively, which are the main decay modes. $\pi_2\rho$, $h_1\eta$, $b_1\pi(1300)$ and $f_2\omega$ are its important decay channels. This work suggests that experimentlist should search for this missing state in $b_1\pi$ final state. Otherwise, $\pi(1300)\rho$, $a_2\rho$, $\pi\rho(1450)$, $a_0\rho$ and $h_1\eta(1295)$ have sizeable contributions to the total width of $\omega_3(3G)$. $\rho_3(3G)$ has the total width of about 270 MeV from our calculation. The main decay modes are {\color{black}$\pi\pi_2$}, $\rho\rho$, $h_1\pi$, $a_1\omega$ and $a_1\pi$, which branching ratios are 0.17, 0.090, 0.089, 0.086 and 0.078. $a_2\pi$, $a_2(1700)\pi$, $f_2\rho$, {\color{black}$\eta(1295)\rho$} and $\rho\rho(1450)$ are the important decay modes of $\rho_3(3G)$.

\begin{table}[ht]
\centering%
\caption{The partial decay widths of the $\omega_3(1G)$, $\rho_3(1G)$,  the unit of width is MeV. The $\gamma$ value is 11.6. \label{decay31G}}
\[\begin{array}{cccc}
\hline
\hline
\multicolumn{2}{c}{\omega_3(2255), \Gamma_{exp.} = 175 \pm $30~\cite{Zyla:2020zbs}$}  &\multicolumn{2}{c}{\rho_3(2255)}\\\hline
\text{Channel}    &\text{Value } &\text{Channel}   &\text{Value}  \\ \midrule[1pt]
\text{Total}     &$325$       &\text{Total}      &$721$\\
b_1\pi            &$147$         &{\color{black}\pi\pi_2}          &$296$\\
{\color{black}\pi\rho_3}         &$47.0$         &b_1\rho           &$76.5$\\
h_1\eta           &$32.9$         &\rho\rho          &$62.6$\\
a_2\rho           &$23.2$         &a_1\pi            &$57.4$\\
{\color{black}\rho\pi}           &$17.4$         &h_1\pi            &$46.2$\\
KK_1^\prime       &$12.5$         &a_2\pi            &$37.1$\\
f_2\omega         &$11.3$         &b_1\eta           &$31.5$\\
\pi\rho(1450)     &$8.69$         &a_1\omega         &$28.9$\\
\pi(1300)\rho     &$5.32$         &{\color{black}\pi\omega_3}       &$16.6$\\
\eta\omega        &$4.00$         &\pi\pi(1300)      &$14.2$\\
\pi\rho(1700)     &$2.57$         &f_2\rho           &$12.2$\\
KK_1              &$2.53$         &{\color{black}K{K_1^\prime}}     &$11.7$\\
KK_2^*            &$2.20$         &{\color{black}\pi\omega}         &$5.86$\\
\eta^\prime\omega &$1.75$         &a_2(1700)\pi      &$5.02$\\
a_1{\rho}         &$1.75$         &\omega(1420)\pi   &$3.01$\\
{\color{black}\eta\omega_3}      &$1.52$         &\pi\pi            &$2.90$\\
\eta\omega(1420)  &$1.05$         &{\color{black}KK_1}              &$2.82$\\
\eta(1295)\omega  &$1.04$         &{\color{black}KK_2^*}            &$2.20$\\
KK^*              &$1.03$         &f_1\rho           &$2.201$\\
                  &              &{\color{black}\eta^\prime\rho}   &$1.87$\\
                  &              &{\color{black}\pi\omega(1650)}   &$1.54$\\
                  &              &b_1\eta^\prime    &$1.27$\\
                  &              &{\color{black}\eta(1295)\rho}   &$1.15$\\
                  &              &{\color{black}KK^*}              &$1.03$\\
 \hline
 \hline
\end{array}\]
\end{table}

\begin{table}[ht]
\centering
\caption{The partial decay widths of the $\omega_3(2G)$, $\rho_3(2G)$,  the unit of width is MeV. The $\gamma$ value is 11.6.      \label{decay32G}}
\[\begin{array}{cccc}
\hline
\hline
\multicolumn{2}{c}{\omega_3 (2520)}  &\multicolumn{2}{c}{\rho_3(2520)}\\
\hline
\text{Channel}    &\text{Value }   &\text{Channel}    &\text{Value}  \\ \midrule[1pt]
\text{Total }     &$312$         &\text{Total}      &$397$        \\
{b_1}\pi          &$107$           &{\color{black}\pi\pi_2}          &$107$\\
a_2\rho           &$40.9$           &h_1\pi            &$37.1$\\
\pi_2\rho         &$21.4$           &a_1\pi            &$36.0$\\
{\color{black}\pi\rho_3}         &$19.4 $          &\rho\rho(1450)    &$31.4$\\
h_1\eta           &$18.9 $          &a_1\omega         &$22.6$\\
\pi\rho           &$18.6$           &b_1\rho           &$21.0$\\
f_2\omega         &$17.5 $          &f_2\rho           &$18.4$\\
KK_2              &$15.3$           &b_1\eta           &$16.4$\\
f_1\omega         &$13.2$           &a_2(1700)\pi      &$16.4$\\
a_0{\rho}         &$9.88$          &a_2\pi            &$16.04$\\
\pi(1300)\rho     &$4.88$           &\rho\rho          &$9.21$\\
\eta_2\omega      &$3.57$           &\pi_2\omega       &$7.03$\\
f_1{h_1}          &$3.03$           &{\color{black}\pi\omega_3}       &$6.81$\\
h_1\eta(1295)     &$3.16$           &b_1{b_1}          &$6.46$\\
f_2{h_1}          &$3.11$           &a_4(2040)\pi      &$6.16$\\
\eta\omega        &$3.08 $          &\pi\pi            &$6.15$\\
f_1(1420)\omega   &$3.03$           &f_1\rho           &$4.72$\\
h_1\eta^\prime    &$2.63$           &h_1\pi(1300)      &$4.35$\\
{\color{black}\eta\omega_3}      &$1.81$           &{\color{black}KK_2^*}            &$3.81$\\
a_1\rho           &$1.31$           &a_2\omega         &$3.50$\\
\eta^\prime\omega &$1.06$           &f_1(1420)\rho     &$3.20$\\
                 &                 &\omega{a_0}       &$3.08$\\
                 &                 &h_1{a_1}          &$2.52$\\
                  &                &a_2h_1            &$2.18$\\
                  &                &{\color{black}\eta\rho_3}        &$1.50$\\
                  &                &KK(1630)          &$1.42$\\
                  &                &\pi\pi(1300)      &$1.38$\\
                  &                &b_1\eta^\prime    &$1.31$\\
 \hline
 \hline
\end{array}\]
\end{table}

\begin{table}[ht]
\centering%
\caption{The partial decay widths of the $\omega_3(3G)$, $\rho_3(3G)$,  the unit of width is MeV. The $\gamma$ value is 11.6.\label{decay33G}}
\[\begin{array}{cccc}
\hline
\hline
\multicolumn{2}{c}{\omega_3(2750)}       &\multicolumn{2}{c}{\rho_3(2750)}\\
\hline
\text{Channel}          &\text{Value }     &\text{Channel}    &\text{Value}  \\ \midrule[1pt]
\text{Total }           &181            &\text{Total}        &270        \\
b_1\pi                 &63.8             &{\color{black}\pi\pi_2}            &46.0\\
\pi\rho                &12.3             &\rho\rho             &24.3\\
{\color{black}\pi\rho_3}              &12.0             &h_1\pi               &24.0\\
f_1\omega              &11.7             &a_1\omega           &23.6\\
\pi_2\rho              &10.4             &a_1\pi              &21.0\\
h_1\eta                &9.84             &a_2\pi               &10.5\\
b_1\pi(1300)           &8.28             &a_2(1700)\pi         &9.40\\
f_2\omega              &8.25             &f_2\rho              &8.45\\
\pi(1300)\rho          &7.77             &{\color{black}\eta(1295)\rho}       &8.17\\
{a_2}\rho              &5.59             &\rho\rho(1450)       &7.91\\
\pi\rho(1450)          &5.32             &b_1\eta              &7.90\\
{a_0}\rho              &4.89             &b_1\rho              &7.84\\
h_1\eta(1295)          &4.21             &a_2(1700)\omega      &7.05\\
{\eta_2}\omega         &1.96             &{h_1}{\pi(1300)}     &6.56\\
{f_1(1420)}\omega      &1.85             &{\pi}{\pi}           &6.40\\
\eta\omega             &1.59             &{\color{black}\pi{\omega_3}}        &4.30\\
{f_2}{h_1}             &1.59             &{\color{black}\pi\omega}            &4.040\\
\eta(1295)\omega       &1.55             &a_4(2040)\pi         &3.94\\
K_1K_1                 &1.50             &\pi_2\omega          &3.55\\
\eta\omega(1420)       &1.50             &{\color{black}\pi(1300)\omega}      &2.62\\
{a_1}\rho              &1.37             &a_1\pi(1300)         &2.17\\
K{K_1^\prime}          &1.34             &f_1(1420)\rho        &1.96\\
{\color{black}\eta\omega_3}           &1.07             &a_1\omega(1420)      &1.86\\
h_1{\eta^\prime}       &1.01             &b_1\eta(1295)        &1.84\\
                       &                  &a_2\omega            &1.69\\
                       &                  &{\color{black}\eta(1475)\rho}       &1.64\\
                       &                  &a_2(1320){h_1}       &1.55\\
                       &                  &b_1{b_1}             &1.54\\
                       &                  &b_1{f_1}            &1.51\\
                       &                  &K_1K_1               &1.50\\
                       &                  &{\color{black}\eta\rho(1450)}       &1.48\\
                       &                  &\omega{a_0}          &1.48\\
                       &                  &\omega(1420)\pi      &1.43\\
                       &                  &h_1{a_1}             &1.39\\
                       &                  &\pi(1300)\pi(1300)   &1.19\\
                       &                  &a_2{a_2}             &1.03\\
 \hline
 \hline
\end{array}\]
\end{table}

\section{CONCLUSION}\label{sec4}

In this work, we have systematically studied the mass spectra and the OZI-allowed two-body strong decay behaviors of the newly observed $X(2220)$ state as well as $\omega$ and $\omega_3$ and $\rho$ and $\rho_3$ states. We adopt a new way to get the value of $\gamma$ in $\omega$ and $\omega_3$ system. Following is the main work.

\begin{enumerate}

\item{The newly observed state $X(2220)$ should be a $3^3D_1$ $\omega$ state.}

\item{The $\omega(782)$, $\omega(1420)$ and $\omega(1960)$ states can be interpreted as the {\color{black}$\omega(1^3S_1)$, $\omega(2^3S_1)$ and $\omega(3^3S_1)$. $\omega(2240)$, $\omega(2330)$ and $\omega(2290)$ can be the candidates of $\omega(4^3S_1)$.}  $\rho(1450)$, $\rho(1900)$ and $\rho(2150)$ are the first, second and third excited states of $\rho(770)$ as shown in Fig. \ref{regge}.}

\item{The mass and width of $\omega(2D)$, by our prediction, are $2047.6$ MeV and  $212 \pm 36.1$ MeV. $\omega(2205)$ and $X(2220)$ are the $\omega(3D)$ states.}

\item{$\omega_3(1945)$, $\omega_3(2285)$ and $\rho_3(1990)$, $\rho_3(2250)$ are the first and second excited states of $\omega_3(1670)$ and $\rho_3(1690)$.}

\item{The masses of $\omega_3(4D)$ and $\rho_3(4D)$ are both predicted to be 2.63 GeV, the width of them are 288 MeV and 420 MeV. We also predict the mass spectra and decay behaviors of $\omega_3(1G)$ and $\rho_3(1G)$, $\omega_3(2G)$ and $\rho_3(2G)$ and $\omega_3(3G)$ and $\rho_3(3G)$.}

\end{enumerate}

 In addition, it is our hope and belief that more experimental measurements will be released in the future.

\section{ACKNOWLEDGE}
This work is supported  by the National Natural Science Foundation of China under Grants No. 11965016 and the Natural Science Foundation of Qinghai Province (2020-ZJ-728). Y. -R. W. and T.-Y. L. contributed equally to this work.

\bibliographystyle{apsrev4-1}
\bibliography{hepref}
\end{document}